\documentclass{article}
\usepackage{epsf}
\newcommand{\eps}[2]{\centering\parbox{#1}{\epsfxsize=#1\epsfbox{#2}}}
\newlength{\picwidth}
\begin{document}
\bibliographystyle{plain}
\title{\bf The sun as a high energy neutrino source}
\author{Christian Hettlage, Karl Mannheim\\Universit\"{a}tssternwarte, Geismarlandstra{\ss}e 11, D-37083 G\"{o}ttingen, Germany\\\hspace*{1mm}\\and\\\hspace*{1mm}\\John G. Learned\\Department of Physics and Astronomy, University of Hawaii,\\Honolulu, Hawaii 96822, USA}
\maketitle

\begin{abstract}
Cosmic ray interactions in the solar atmosphere yield a flux of electron and muon neutrinos with energies greater than 10 GeV. We discuss the influence of neutrino oscillations on the event rates in water-based \v{C}erenkov detectors due to this neutrino flux and comment on the possibility of detecting the sun as a high energy neutrino source.
\end{abstract}
\section{Introduction}
Cosmic ray impingement on the solar atmosphere leads to the production of secondary particles via high energy $pp$-interactions, the decay of which subsequently results in a flux of both electron and muon neutrinos. (Note that throughout this article the term neutrino is meant to refer to both neutrino and corresponding antineutrino.)

\begin{figure*}
\eps{\picwidth}{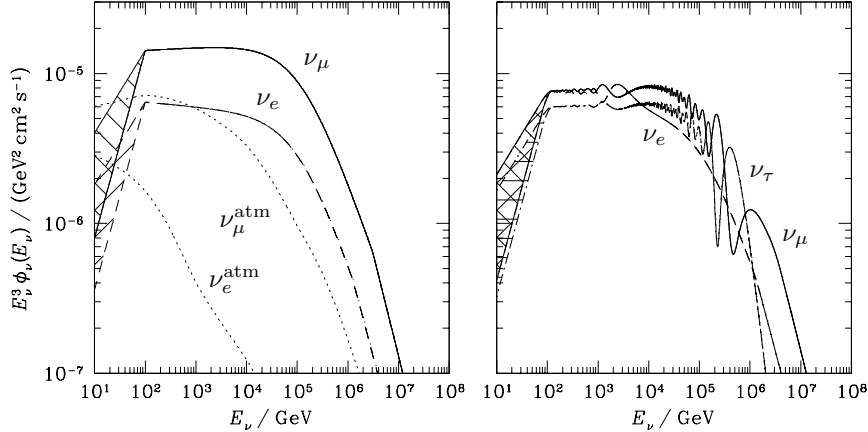}
\put(-0.712205,0.13){$\nu_{e}$}
\put(-0.659904,0.175){$\nu_{\mu}$}
\put(-0.765,-0.071288){$\nu_{e}^{\rm atm}$}
\put(-0.753,-0.005050){$\nu_{\mu}^{\rm atm}$}
\put(-0.284474,0.09){$\nu_{e}$}
\put(-0.1295,-0.020336){$\nu_{\mu}$}
\put(-0.166,0.05 ){$\nu_{\tau}$}
\caption{Fluxes of solar atmosphere neutrinos (at the earth) for $\nu_{e}$ (dashed), $\nu_{\mu}$ (solid), and $\nu_{\tau}$ (dot-dashed line), integrated over the solid angle of the sun. \emph{Left:} The fluxes without neutrino oscillations, as obtained in~\cite{IT96}. For energies lower than 100\,GeV $\phi_{\nu}\propto E_{\nu}^{-\gamma}$ is assumed, and the range from $\gamma=1.75$ to 2.45 is shown. In addition, the plot includes the terrestrial atmospheric horizontal $\nu_{e}$ (lower dotted) and $\nu_{\mu}$ flux (upper dotted line), also integrated over the solar disk~\cite{Vol81}. \emph{Right:} The corresponding fluxes for the choice $\Delta m_{\rm sun}^{2}=1.9\times10^{-5}\,{\rm eV}^{2}$, $\Delta m_{\rm atm}^{2}=3.5\times10^{-3}\,{\rm eV}^{2}$, $\sin\theta_{\rm sun}=0.58$, and $\sin\theta_{\rm atm}=0.86$, averaged over the interval from $10^{-0.1}E_{\nu}$ to $10^{0.1}E_{\nu}$. This averaging is justified by the limited energy resolution of neutrino telescopes, and it smears out rapid neutrino oscillations.}
\label{fig:spectrum}
\end{figure*}

In order to compute the neutrino flux due to these processes, one first has to evaluate the absorption rate of cosmic rays in the sun, taking into account the interplanetary and solar magnetic fields~\cite{SSG91}. The high energy interactions may then be treated by means of Monte Carlo simulations such as JETSET and PYTHIA. Finally, the shadowing effect of inelastic neutrino scattering in the sun has to be included. This analysis is carried out for energies exceeding 100\,GeV in~\cite{IT96}. The results are shown in the left part of Fig.~\ref{fig:spectrum} and will be used in this article. For energies $E_{\nu}$ lower than 100\,GeV we assume that the flux $\phi_{\nu_{e/\mu}}$ of solar atmosphere neutrinos integrated over the solar disk is given by $\phi_{\nu_{e/\mu}}\propto E_{\nu_{e/\mu}}^{-\gamma}$, where $1.75<\gamma<2.45$, thus allowing for some uncertainty due to heliomagnetic effects. Seckel et al.~\cite{SSG91} favor the lower limit of $\gamma$. Note that both choices are consistent with the EGRET limit on the gamma ray flux of the quiet sun~\cite{Tho97}, if a smaller value of $\gamma$ is adopted for $E_{\nu}<10\,{\rm GeV}$.

The solar atmosphere neutrino spectra thus obtained may be altered by neutrino oscillations, which depend on the neutrino mass differences and mixing matrices. Together with the suggested solutions to the solar neutrino problem and long baseline experiments, the SuperKamiokande data on terrestrial atmospheric neutrinos can be used to narrow down these parameters to a few cases.

In this article we investigate, for the various mixing matrices and corresponding mass differences, the influence of neutrino oscillations on the expected event rate of solar atmosphere neutrinos.

\section{Neutrino detection}
For the detection of the solar atmosphere neutrino flux water-based \v{C}erenkov detectors may be used. As these detect the \v{C}erenkov radiation of leptons produced in charged current interactions, the total event rate $\dot{N_{\nu}}$ for energies exceeding some threshold $E_{0}$ is approximately given by
\begin{equation}
\dot{N}_{\nu}=\int_{E_{0}}^{\infty}{\rm d}E\,\phi_{\nu}(E)\,\sigma_{\rm CC}(E)\,\frac{\rho}{m_{p}}\,L_{\nu}(E)\,A,
\label{eq:eventrate}
\end{equation}
where $m_{p}$ denotes the proton mass, $\rho=1\,{\rm g/cm}^{3}$ the matter density, and $A$ the effective detector area. For computing the charged current cross section $\sigma_{\rm CC}$ the CTEQ4DIS parton distribution functions~\cite{Gan98} are used. Concerning tau neutrinos, the phase space limitations due to the large tauon mass must be taken into account. $L_{\nu}$ is the lepton range or the detector thickness $h$ in the direction of the sun, whichever the larger. Hence for $\nu_{e}$ and $\nu_{\tau}$ we may assume $L_{\nu}=h$, whereas for $\nu_{\mu}$ the relation
\[L_{\mu}(E)=\max\left\{\frac{1}{\beta\rho}\ln\frac{E+\alpha/\beta}{E_{0}+\alpha/\beta},\,h\right\}\]
with $\alpha=2.5\,{\rm MeV/(g\,cm^{-2})},\ \beta=4.0\times10^{-6}\,{\rm (g\,cm^{-2})^{-1}}$~\cite{IT96} has to be employed. In the following, we will assume $E_{0}=10\,{\rm GeV}$. Taking $A=10^{4}\,{\rm m}^{2}$, $h=500\,{\rm m}$ and $A=1\,{\rm km}^{2}$, $h=1\,{\rm km}$ as an example, the solar atmosphere neutrino fluxes of Fig.~1 yield event rates of $\dot{N}_{e}=0.1\,\mbox{--}\,0.2\,{\rm a}^{-1}$, $\dot{N}_{\mu}=0.3\,\mbox{--}\,0.5\,{\rm a}^{-1}$ and $\dot{N}_{\rm e}=24\,\mbox{--}\,46\,{\rm a}^{-1}$, $\dot{N}_{\mu}=46\,\mbox{--}\,82\,{\rm a}^{-1}$, respectively. The range of values reflects the allowed range of $\gamma$ for energies smaller than 100\,GeV.

\section{Neutrino oscillations}
\begin{figure}
\eps{8.8cm}{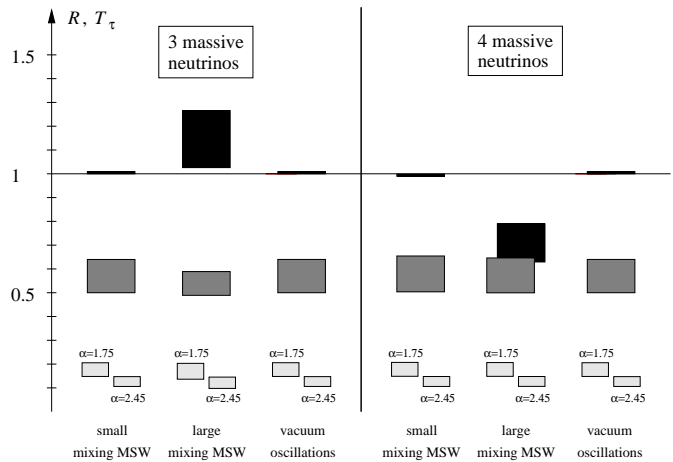}
\caption{$R_{e}$ (black bars), $R_{\mu}$ (dark grey bars), and $T_{\tau}$ (light grey bars) for the various neutrino mixing schemes, if $E_{0}=10\,{\rm GeV}$ is assumed. $R_{e}$ and $R_{\mu}$ are given for $\gamma=1.75$, $T_{\tau}$ for $\gamma=1.75$ and $\gamma=2.45$. The bars show the range of values allowed by the uncertainty of the mixing angle values and the neutrino masses. In case of no neutrino oscillations, $R_{e/\mu}$ and $T_{\tau}$ would be given by $R_{e/\mu}=1$ and $T_{\tau}=0$.}
\label{fig:RT}
\end{figure}

So far, no neutrino oscillations have been taken into account. However, the solar neutrino problem~\cite{Hax95} and the SuperKamiokande atmospheric neutrino data~\cite{Fuk98} are best explained by transitions between the various neutrino flavors, so that one should expect that the solar neutrinos oscillate on their flight from sun to earth. Then the probability of a solar neutrino of flavor $\alpha$ arriving as a neutrino of flavor $\beta$ is given by~\cite{BGG98}
\[P_{\nu_{\alpha}\longrightarrow\nu_{\beta}}=\left|\delta_{\alpha\beta}+\sum_{k=2}^{n}U_{\beta k}U_{\alpha k}^{\ast}\left[\exp\left(-i\frac{\Delta m_{k1}^{2}L}{2E_{\nu}}\right)-1\right]\right|^{2}\]
with the distance $L$ between earth and sun, $\Delta m_{k1}^{2}\equiv m_{k}^{2}-m_{1}^{2}$ ($m$ being the neutrino mass), and the mixing matrix $U$. Note that because of the high energies Mikheyev-Smirnov-Wolfenstein (MSW) effects may be ignored. The experimental data on neutrino oscillations suggest mixing matrices of the form~\cite{BGG98}
\[U=\left(\begin{array}{ccc}
  \cos\theta_{\rm sun} & \sin\theta_{\rm sun} & 0\\
  -\sin\theta_{\rm sun}\cos\theta_{\rm atm} & \cos\theta_{\rm sun}\cos\theta_{\rm atm} & \sin\theta_{\rm atm}\\
  \sin\theta_{\rm sun}\sin\theta_{\rm atm} & -\cos\theta_{\rm sun}\sin\theta_{\rm atm} & \cos\theta_{\rm atm}
          \end{array}\right)\]
for 3 flavors and
\[U=\left(\begin{array}{cccc}
  0 & 0 & \cos\theta_{\rm sun} & \sin\theta_{\rm sun}\\
  \cos\theta_{\rm atm} & \sin\theta_{\rm atm} & 0 &0\\
  -\sin\theta_{\rm atm} & \cos\theta_{\rm atm} & 0 & 0\\
  0 & 0 & -\sin\theta_{\rm sun} & \cos\theta_{\rm sun}
          \end{array}\right)\ {\rm (case\ A)}\]
or
\[U=\left(\begin{array}{cccc}
  \cos\theta_{\rm sun} & \sin\theta_{\rm sun} & 0 & 0\\
  0 & 0 & \cos\theta_{\rm atm} & \sin\theta_{\rm atm}\\
  0 & 0 & -\sin\theta_{\rm atm} & \cos\theta_{\rm atm}\\
  -\sin\theta_{\rm sun} & \cos\theta_{\rm sun} & 0 & 0
          \end{array}\right)\ {\rm (case\ B)}\]
for 4 flavors, i.e. if the existence of a sterile neutrino is assumed. In the following discussion the two 4 flavor matrices lead to the same results, as the schemes of case A and B both consist of a $\nu_{e}$-$\nu_{\rm sterile}$ oscillation, which is relevant for the solar neutrino problem, and a $\nu_{\mu}$-$\nu_{\tau}$ oscillation, which is relevant for the atmospheric neutrino data. The experimental limits of the mixing angles $\theta_{\rm atm}$ and $\theta_{\rm sun}$ and of the mass square differences $\Delta m_{k1}^{2}$ for the various solutions of the solar neutrino problem are listed in Table~\ref{tab:parameters}.
\begin{table*}
\begin{tabular}{lll}
mixing scheme & mixing angle & mass square difference\\[1ex]\hline\\[-1.8ex]
small mixing MSW & $0.003\leq\sin^{2}2\theta_{\rm sun}\leq0.011$ & $4\times10^{-6}\,{\rm eV}^{2}\leq\Delta m_{21}^{2}\leq1.2\times10^{-5}\,{\rm eV}^{2}$\\[1ex]
large mixing MSW & $0.42\leq\sin^{2}2\theta_{\rm sun}\leq0.74$ & $8\times10^{-6}\,{\rm eV}^{2}\leq\Delta m_{21}^{2}\leq3.0\times10^{-5}\,{\rm eV}^{2}$\\[1ex]
vacuum oscillations & $0.70\leq\sin^{2}2\theta_{\rm sun}\leq1$ & $6\times10^{-11}\,{\rm eV}^{2}\leq\Delta m_{21}^{2}\leq1.1\times10^{-10}\,{\rm eV}^{2}$\\[1ex]\hline\\[-1.8ex]
\emph{for all mixing schemes:} & $0.72\leq\sin^{2}2\theta_{\rm atm}\leq1$ & $4\times10^{-4}\,{\rm eV}^{2}\leq\Delta m_{31}^{2}\leq8\times10^{-3}\,{\rm eV}^{2}$
\end{tabular}
\caption{Mixing angles $\theta_{\rm atm}$ and $\theta_{\rm sun}$ and mass square differences $\Delta m_{k1}^{2}$ for the small mixing angle MSW, the large mixing angle MSW, and the vacuum oscillation solution of the solar neutrino problem, where 3 neutrino flavors are assumed. For 4 neutrino flavors the substitutions $\Delta m_{21}^{2}\longrightarrow\Delta m_{43}^{2}$, $\Delta m_{31}^{2}\longrightarrow\Delta m_{21}^{2}$ (case A) or $\Delta m_{31}^{2}\longrightarrow\Delta m_{43}^{2}$ (case B) must be made. The limits are taken from~\cite{BGG98}.}
\label{tab:parameters}
\end{table*}
The differential fluxes $\phi_{\nu}$ of $\nu_{e}$, $\nu_{\mu}$, and $\nu_{\tau}$ at the earth may now be written as
\begin{eqnarray*}
\phi_{e} & = & \phi_{e,\rm sun}P_{\nu_{e}\longrightarrow\nu_{e}}+\phi_{\mu\rm,sun}P_{\nu_{\mu}\longrightarrow\nu_{e}}\\
\phi_{\mu} & = & \phi_{e,\rm sun}P_{\nu_{e}\longrightarrow\nu_{\mu}}+\phi_{\mu\rm,sun}P_{\nu_{\mu}\longrightarrow\nu_{\mu}}\\
\phi_{\tau} & = & \phi_{e,\rm sun}P_{\nu_{e}\longrightarrow\nu_{\tau}}+\phi_{\mu,\rm sun}P_{\nu_{\mu}\longrightarrow\nu_{\tau}}.
\end{eqnarray*}
The right part of Fig.~\ref{fig:spectrum} shows an example of solar atmosphere neutrino fluxes with neutrino oscillations. One obtains the corresponding total event rates by inserting the fluxes at the earth into Eq.~\ref{eq:eventrate}. In order to compare the rates thus obtained to those without neutrino oscillations, we introduce the ratios
\[R_{e/\mu}\equiv\frac{\mbox{total $\nu_{e/\mu}$ event rate with neutrino oscillations}}{\mbox{total $\nu_{e/\mu}$ event rate without neutrino oscillations}}\]
\[T_{\tau}\equiv\frac{\mbox{total $\nu_{\tau}$ event rate with neutrino oscillations}}{\mbox{total $\nu_{\mu}$ event rate without neutrino oscillations}}.\]
Clearly, $R_{e/\mu}$ and $T_{\tau}$ depend on $U$, $L$ and $\Delta m_{k1}^{2}$. However, due to the fact that these quantities involve an integration over the energy, the dependene on $L$ and $\Delta m_{k1}^{2}$ within the mass square ranges given in Table~\ref{tab:parameters} is weak and can be neglected. Furthermore, there is virtually no dependence on the detector size. The ranges of $R_{e/\mu}$ and $T_{\tau}$ for the various mixing schemes are given in Fig.~\ref{fig:RT}. $R_{e}$ and $R_{\mu}$ essentially do not depend on the precise value of $\gamma$ between 1.75 and 2.45. However, because of the limited tauon phase space this is not true for $T_{\tau}$.

\section{Background}
Concerning the solar atmosphere neutrino flux there are three possible kinds of background fluxes:
\begin{itemize}
\item For electron and muon neutrinos cosmic ray impingement on earth's atmosphere leads to a background flux (cf. Fig.~\ref{fig:spectrum}), which in both cases is of the same order as the solar atmosphere neutrino flux. For tau neutrinos the terrestrial atmospheric flux can be neglected for energies far above 1~GeV. Therefore in all three cases, the solar flux will be discernible from the terrestrial atmospheric background.
\item The decay of WIMPs in the solar interior might produce a neutrino flux exceeding the one due to cosmic ray interactions~\cite{BEG98}.
\item Blazars, gamma-ray bursts, and particle decays may give rise to an isotropic neutrino background~\cite{LM99}. Assuming that an upper limit to this background flux is given by $\phi_{\nu}(E)=\xi\,k\,(E/1\,{\rm GeV})^{-\kappa}$ with the coefficients $\kappa=2.1$, $k=7.32\times10^{-6}\,{\rm cm^{-2}s^{-1}sr^{-1}GeV^{-2}}$ of the extragalactic gamma-ray spectrum as obtained by EGRET~\cite{Sre98} and $\xi\leq1$, one obtains even for a cubic kilometer telescope an event rate less than one event per year. Shadowing and cascading of neutrinos in the sun only further diminish this result. Hence the isotropic background can safely be neglected.
\end{itemize}
Accordingly, the background is lower than the solar neutrino flux. It might be, however, that the solar neutrino flux is dominated by neutrinos due to the decay of WIMPs. The impact of the uncertainty in the knowledge of the initial neutrino direction will be discussed in the next section.

\section{Discussion}
\begin{figure}
\eps{8.8cm}{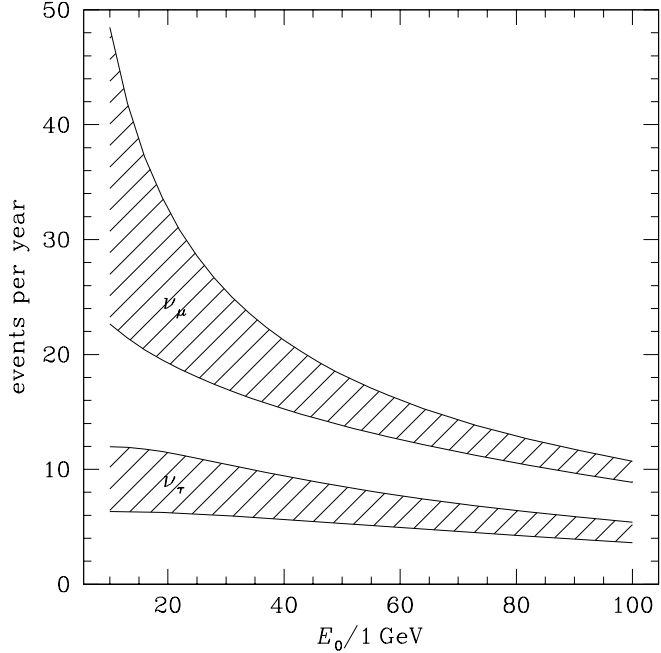}
\caption{$\nu_{\mu}$ and $\nu_{\tau}$ event rates in a 1\,km$^{3}$ detector as a function of the threshold energy $E_{0}$ for the large mixing MSW case with three neutrino flavors, taking into account the uncertainty of the initial solar neutrino spectra, the mixing angles and the neutrino masses.}\label{fig:threshold}
\end{figure}

From Fig.~\ref{fig:RT} we see immediately that the influence of neutrino oscillations is mostly independent of the mixing scheme, the only exception being the large mixing angle MSW case for electron neutrinos. Assuming a threshold energy of $E_{0}=10\,{\rm GeV}$, one may predict the event rates in water-based \v{C}erenkov detectors with an effective area of $A=10^{4}\,{\rm m}^{2}$ ($A=1\,{\rm km}^{2}$) and a thickness of $h=500\,{\rm m}$ ($h=1\,{\rm km}$) to be $0.08\,\mbox{--}\,0.3\,{\rm a}^{-1}$ ($12\,\mbox{--}\,44\,{\rm a}^{-1}$) for electron, $0.1\,\mbox{--}\,0.3\,{\rm a}^{-1}$ ($23\,\mbox{--}\,53\,{\rm a}^{-1}$) for muon and $0.03\,\mbox{--}\,0.1\,{\rm a}^{-1}$ ($6\,\mbox{--}\,12\,{\rm a}^{-1}$) for tau neutrinos. The corresponding $\nu_{\mu}$ and $\nu_{\tau}$ event rates in a 1\,km$^{3}$ detector for threshold energies greater than 10\,GeV are given in Fig.~\ref{fig:threshold} for the large mixing MSW case with three neutrino flavors. For comparison, we note that the rate of tau neutrino events to be expected from the proposed CNGS beam is of the order of 30 per year~\cite{Bai99,Ell99}.

 Hence at first sight it seems that, although not producing a sufficiently high event rate in present-day detectors, the sun should be detectible with next-generation neutrino telescopes. However, there are three drawbacks: Firstly, due to the immense number of atmospheric high energy muons, for $\nu_{\mu}$ only upward-going neutrinos (i.e. with zenith angles greater or (depending on the detector depth) slightly less than $90^{\circ}$) may be detected. For the sun, this reduces the $\nu_{\mu}$ event rate to about half its value, the precise amount depending on the detector depth and latitude.

Secondly, at present the detection rate of unambigous neutrino events is lower than the actual event rate (cf.~\cite{And99}).

Thirdly, for the energy range considered in this article the mean angle between neutrino and corresponding lepton cannot be neglected. For muon neutrinos it is given by $1.5^{\circ}\,(E_{\nu}/100\,{\rm GeV})^{-0.5}$. Hence effectively the solid angle of the sun is enlarged, so that the terrestrial background exceeds the solar flux by up to three orders of magnitude for $\nu_{\rm e}$ and $\nu_{\mu}$. 

The low angular resolution could be improved if information on the hadronic cascade might be used. Alternatively, one may restrict the neutrino energies to values greater than 100\,GeV. In this case the number of solar atmosphere muon neutrino events for a horizontal flux and a one-year run (i.e. 7--10 events) would be comparable to the statistical error of the number of background events. 

It should be noted that due to the lack of any significant background, for energies far above 1 GeV the tau neutrino detectibility is not affected by the solid angle over which one has to integrate.

Evidently, the detection of the sun requires intelligent event reconstruction schemes combined with very fast read-out detectors. Concerning tauons, the ``double bang'', i.e. the two cascades arising at the production and the decay of a tauon~\cite{LP95}, may be used, if a sufficiently fine granularity of the neutrino detector is achieved. The sun might thus serve as a test beam for searching neutrino oscillations on a scale exceeding earthbound distances. In particular, it offers the prospect of the discovery of tau-appearance.

\section*{Acknowledgements}
We thank F. Rieger and D. Horns for helpful discussions. Part of this work has been supported by the Studienstiftung des deutschen Volkes.
\bibliography{literature}
\end{document}